\documentclass{aastex631}
\usepackage{subfiles}
\usepackage{array}
\usepackage{amsmath}
\usepackage{booktabs}
\usepackage{pifont}

\begin{document}

\title{Shaping the Mantle: The Role of Superheated Core After Giant Impacts}

\correspondingauthor{You Zhou}
\email{youzhou.sim@protonmail.com}


\author[0000-0002-0786-7307]{You Zhou}
\affiliation{College of Earth and Planetary Sciences, Chengdu University of Technology, \\ Chengdu 610059, China}
\affiliation{CAS Center for Excellence in Comparative Planetology, Hefei 230026, China}







\begin{abstract}

The Moon-forming giant impact significantly influenced the initial thermal state of Earth's mantle by generating a global magma ocean, marking the onset of mantle evolution. Recent Smoothed Particle Hydrodynamics (SPH) simulations indicate that such a collision would produce a superheated core, and its cooling should strongly influence subsequent mantle dynamics. Here, we present systematic SPH simulations of diverse giant-impact scenarios and show that the superheated core formed after the impact can trigger secondary mantle melting, thereby governing the final state of the magma ocean. To further quantify this effect, we employed a parameterized core-cooling model to evaluate the influence of secondary melting on the lower mantle. Our results suggest three possible outcomes: complete mantle melting, the formation of a basal melt layer, or the initiation of an early superplume. Combined with recent two-phase magma ocean solidification models, we infer that all three scenarios would result in basal melt layers of varying thickness, which would partially retain the thermal energy of the superheated core.
In the canonical Moon-forming scenario, the superheated core would rapidly transfer heat to Earth’s lower mantle, causing secondary mantle melting within approximately 277--5983~years
, and generating either a basal melt layer or a fully molten mantle. Both outcomes would effectively erase primordial heterogeneities in the lower mantle and impose distinct pathways for the subsequent thermal evolution of the mantle.


\end{abstract}




\section{Introduction} \label{sec:1}

For decades, the Giant Impact Hypothesis has been the dominant theory for explaining the Moon’s origin, as it provides a comprehensive explanation for key dynamical and geochemical features of the Earth–Moon system (e.g., \citep{canup2001origin,canup2012forming,cuk2012making,reufer2012hit,rufu2017multiple,hosono2019terrestrial,kegerreis2022immediate}). This cataclysmic collision not only formed the Moon but also established the initial thermal, physical, and chemical states of the proto-Earth \citep{nakajima2015melting,nakajima2021scaling,zhou2024scaling}, thereby shaping Earth’s subsequent evolutionary path. As the most energetic event during Earth’s accretion, the giant impact would have melted a large portion of the mantle, naturally leading to the formation of a global magma ocean. 

However, the giant impact not only melted the mantle but could also have directly heated the core, producing a superheated interior. Earlier studies have proposed that impact-induced shock heating could create thermally stratified layers in planetary cores, thereby suppressing convection and shutting down dynamo action \citep{arkanihamed2010,roberts2014impact}, or alternatively disrupting stratification and triggering dynamo onset \citep{reese2010,jacobson2017}. However, these studies did not employ SPH simulations and likely underestimated the extent of core heating. Using SPH simulations, \citet{carter2020energy} and \citet{lock2020energy} demonstrated that impact energy is deposited in a highly heterogeneous manner, with complex exchanges among kinetic, potential, and internal energy reservoirs, ultimately leading to localized vaporization, thermal stratification, and prolonged post-impact recovery. \citet{nakajima2021scaling} developed mantle melt scaling relations to precisely characterize mantle heat distribution and highlighted significant pressure contrasts between global magma oceans and localized melt pools. \citet{marchi2023long} simulated the post-impact thermal structure of Venus and proposed that shock heating during late accretion could have significantly raised the temperature of Venus’s core, profoundly affecting its long-term thermal evolution and surface volcanism. Building on these results, \citet{zhou2024scaling} systematically modeled impact-induced core heating and subsequent core cooling, concluding that core superheating is an inevitable consequence of giant impacts. Specifically, a canonical Moon-forming collision could raise core–mantle boundary temperatures to ~11,000 K, thereby delaying the onset of the geodynamo by ~250 million years.

Recent SPH studies have also shown that impact-induced superheated cores contain substantially more thermal energy than previously thought, which would promote the development of a stable thermal stratification on the core side of the core–mantle boundary \citep{zhou2024scaling,marchi2023long}. As a result, heat from the superheated region would be rapidly transferred primarily upward into the mantle, rather than conducted into the deeper core. This upward heat flux would inevitably trigger secondary mantle melting and shape the mantle’s subsequent thermal evolution.

To address these issues, we employ SPH simulations to investigate the initial mantle melting state produced by different giant impacts. We then couple the effects of core superheating with parameterized thermal models to evaluate the extent of secondary mantle melting. Finally, drawing on recent two-phase flow simulations, we infer the possible ultimate fate of the basal melt layer. Our results suggest that, in the short term, core superheating and secondary melting would disrupt the primordial heterogeneity of the lower mantle, while in the long term, they would give rise to a basal magma ocean of varying scale. Overall, these findings demonstrate that a superheated core governs the mantle's final thermal structure, inevitably produces a basal melt layer, and thereby exerts a persistent influence on Earth’s thermal evolution.

\section{Methods} \label{sec:2}

\subsection{SPH Model} \label{sec:2.1}

The Smoothed Particle Hydrodynamics (SPH) method is widely used to simulate giant impacts on terrestrial planets. Therefore, we employed a well-established density-corrected SPH code to investigate the formation of superheated cores resulting from different types of collisions (e.g., \citep{zhou2024scaling,reinhardt2020bifurcation,reinhardt2022forming}). The fundamental governing equations used are:

\begin{equation}
\rho_i = \sum_{j=1}^{N} m_j\, W_{ij}
\end{equation}

\begin{equation}
    \frac{D\mathbf{v}_i}{Dt} = -\sum_{j=1}^{N} m_j \left(\frac{P_j}{\rho_j^2} + \frac{P_i}{\rho_i^2}\right) \nabla_i W_{ij},
\end{equation}
\begin{equation}
    \frac{De_i}{Dt} = \frac{1}{2} \sum_{j=1}^{N} m_j \left(\frac{P_j}{\rho_j^2} + \frac{P_i}{\rho_i^2}\right) \mathbf{v}_{ij} \cdot \nabla_i W_{ij},
\end{equation}
where $\rho_i$, $\mathbf{v}_i$, and $e_i$ denote the density, velocity, and internal energy of particle $i$, respectively. $W_{ij}$ is the smoothing kernel with gradient $\nabla_i W_{ij}$, $h$ is the smoothing length that defines particle influence radius, $m_j$ represents neighbor particle mass, and $N$ is the number of neighboring particles. Pressure $P_i$ is computed through an equation of state (EoS): $P_i = \text{EoS}(\rho_i, e_i)$.

To accurately represent discontinuities caused by shocks, we incorporate artificial viscosity terms into momentum and energy equations as follows:
\begin{equation}
    \left. \frac{D\mathbf{v}_i}{Dt} \right|_{\text{av}} = -\sum_{j=1}^{N} m_j \Pi_{ij} \nabla_i W_{ij},
    \label{eq:artificial_visc_momentum}
\end{equation}
\begin{equation}
    \left. \frac{De_i}{Dt} \right|_{\text{av}} = \frac{1}{2} \sum_{j=1}^{N} m_j \Pi_{ij} \mathbf{v}_{ij} \cdot \nabla_i W_{ij},
    \label{eq:artificial_visc_energy}
\end{equation}
where the artificial viscosity $\Pi_{ij}$ is defined as:
\begin{equation}
\Pi_{ij} = \begin{cases}
    \dfrac{-\alpha c_{ij}\mu_{ij} - \beta \mu_{ij}^2}{\rho_{ij}}, & \text{if } (\mathbf{v}_i - \mathbf{v}_j)\cdot(\mathbf{r}_i - \mathbf{r}_j)<0,\\[8pt]
    0, & \text{otherwise,}
\end{cases}
\label{eq:artificial_visc_definition}
\end{equation}
with
\begin{equation}
\mu_{ij} = \frac{h(\mathbf{v}_i - \mathbf{v}_j)\cdot(\mathbf{r}_i - \mathbf{r}_j)}{(\mathbf{r}_i - \mathbf{r}_j)^2 + \eta h^2}.
\label{eq:mu}
\end{equation}
Here, $c_{ij}$ and $\rho_{ij}$ represent the averaged sound speed and density between particles $i$ and $j$, respectively; $\alpha=1.5$, $\beta=3$, and $\eta=0.01$ are parameters following previous studies \citep{canup2004simulations}. $\mathbf{r}_i$ is the position vector of particle $i$.

 Prior to collision simulations, both the proto-Earth and impactor were relaxed to a hydrostatic equilibrium state. A state-of-the-art equation of state (EoS) incorporating a liquid phase was adopted to ensure high-pressure and high-temperature accuracy. Specifically, the Forsterite EoS was used for the mantle, and the Iron EoS for the core \citep{stewart_sarah_t_2019,stewart2020shock}. Both the targets and impactors were composed of approximately 30\% core and 70\% mantle by mass, consistent with previous studies \citep{zhou2021core,zhou2024scaling}.

 After each impact, we recalculated the proto-Earth core's center of mass to determine the core temperature profile, removing particles below 1\% average density to minimize interference. Core-mantle overlaps post-impact were assigned to the core if metal particle proportion exceeded 90\%. Our simulations spanned various impact angles ($15^\circ$, $30^\circ$, $45^\circ$, $60^\circ$, $75^\circ$), velocities ($1V_{\text{esc}}$, $2V_{\text{esc}}$, $3V_{\text{esc}}$), and impactor masses ($0.05M_{\oplus}$, $0.1M_{\oplus}$, $0.2M_{\oplus}$, $0.3M_{\oplus}$), where $V_{\text{esc}}$ is the mutual escape velocity and $M_{\oplus}$ is Earth's mass. In addition to these systematic SPH simulations, we also performed separate simulations for the canonical collision and the sub-Earth collision. All collision details are provided in Table \ref{tab:summary}.

\subsection{Parameterized Mantle Melting Model} \label{sec:2.2}

\begin{figure}[ht!]
\centering
\includegraphics[scale=0.6]{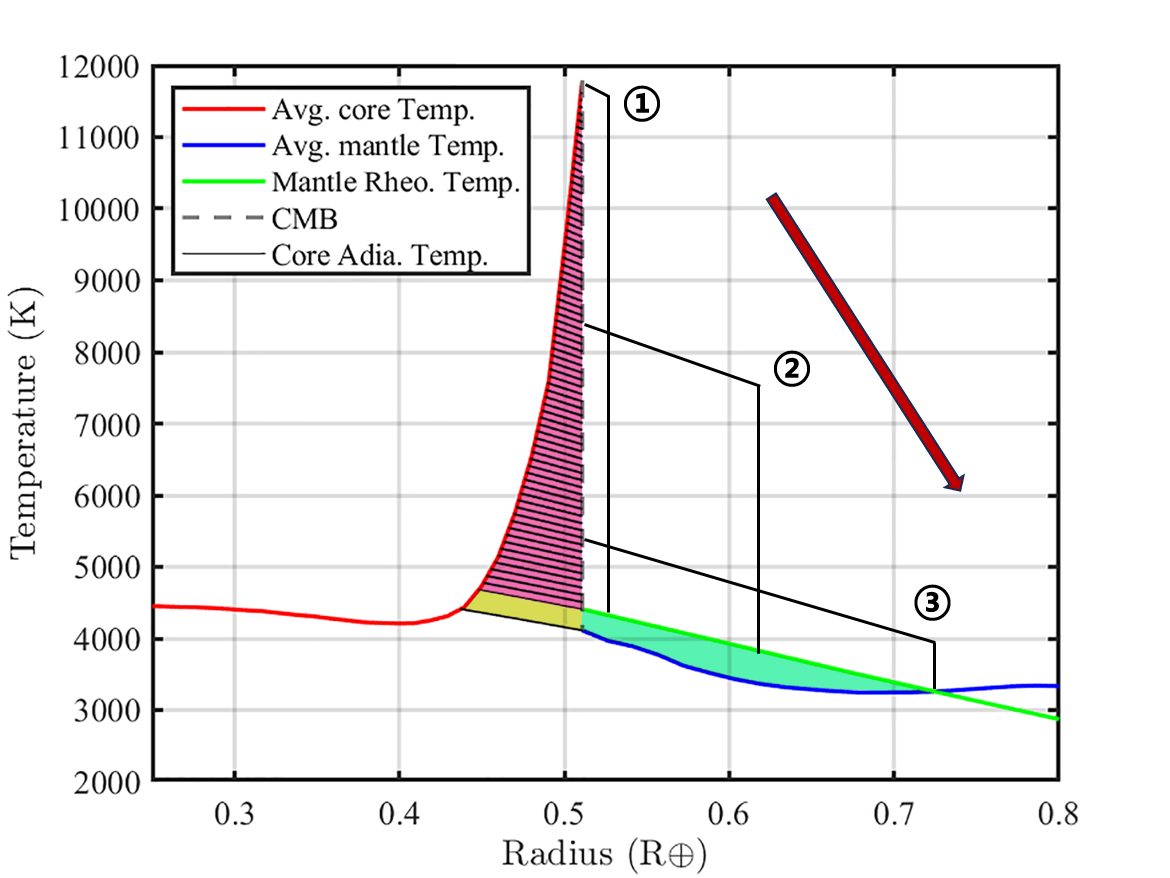}
\caption{\textbf{Schematic illustration of the parameterized mantle melting model.} The left side shows the high-temperature core region, represented by red and yellow blocks, whereas the right side illustrates the cooler, solid mantle in green. Gray lines within the core denote the computational layers used to calculate internal heat transfer, each corresponding to a fixed temperature decrement along the core’s adiabatic profile at the core–mantle boundary (CMB). Black lines on the mantle side depict the progressive expansion of the basal mantle melting layer. The green line represents the rheological transition temperature, which was derived from the solidus and liquidus curves adopted in \citep{monteux2016}. Markers 1 through 3 indicate different stages in the evolution of this melt layer, characterized by decreasing temperature and increasing thickness. Red arrows show the direction and progression of the upward and outward growth of the basal melting layer over time. This schematic conceptually links the core’s thermal structure to the dynamic melting response of the overlying mantle.
\label{fig:1}}
\end{figure}

Essentially, the superheated core would rapidly generate a melting layer at the base of the mantle, with the mantle melting rate primarily controlled by the viscosity of the melt layer \citep{Tackley2025,ke2009coupled}, rather than by that of the lower mantle. Therefore, based on \cite{ke2009coupled} and \cite{driscoll2014thermal}, we developed a parameterized core cooling model specifically adapted to simulate heat transfer from the superheated core to the lower mantle, taking into account our unique thermal structure. This core cooling model can effectively account for a wide range of core-to-mantle temperature gradients. Importantly, it explicitly incorporates the basal melt layer(BML) at the core–mantle boundary, which plays a critical role in regulating thermal flux. Figure~\ref{fig:1} illustrates a schematic of this process.


First, our model computes the total heat content available in the superheated core (corresponding to the regions shown in red and yellow in Figure~\ref{fig:1}), as well as the total heat required to melt the solid mantle into rheological transition temperature (represented by the green region). The formulations used to calculate these heat contents are provided in Equations~\ref{equation8}–\ref{equation11}. Among these, the heat transfer from the superheated core is calculated using the following expression:

\begin{equation}
Q_c = \sum_{i=1}^n C_c M_{ci} (T_{ci} - T_{rheo})
\label{equation8}
\end{equation}

\begin{equation}
M_{ci} = \frac{4}{3} \pi \rho_{ci} (R_{ci}^3 - R_{ci-1}^3)
\label{equation9}
\end{equation}

where $Q_c$ is the heat content transferred from the superheated core, $C_c$ is the specific heat capacity of the core, $M_{ci}$ represents the mass of the material in the $i$-th shell of the core, $T_{ci}$ is the temperature of the $i$-th shell of the core, $T_{rheo}$ is the rheological transition temperature at the core side of the core–mantle boundary (CMB), $\rho_{ci}$ is the density of the $i$-th shell of the core, and $R_{ci}$ and $R_{ci-1}$ are the outer and inner radii of the $i$-th shell of the core, respectively. The expression for calculating the heat content required for the mantle to reach rheological transition temperature is given as follows:

\begin{equation}
Q_m = \sum_{i=1}^n \left[ C_m M_{mi} (T_{\text{rheo}i} - T_{mi}) + \phi_{\mathrm{BML},i} L_i M_{mi} \right]
\label{equation10}
\end{equation}

\begin{equation}
M_{mi} = \frac{4}{3} \pi \rho_{mi} (R_{mi}^3 - R_{mi-1}^3)
\label{equation11}
\end{equation}

where $Q_m$ is the heat content required for the mantle to reach rheological transition temperature, $C_m$ is the specific heat capacity of the mantle, $M_{mi}$ is the mass of the $i$-th mantle shell, $T_{\text{rheo}i}$ is the rheological transition temperature of the $i$-th shell, $T_{mi}$ is its temperature, $\rho_{mi}$ is its density, and $R_{mi}$ and $R_{mi-1}$ are the outer and inner radii of the $i$-th mantle shell, respectively. The term $\phi_{\mathrm{BML},i} L_i$ represents the latent heat required for partial melting in the basal melt layer, where $\phi_{\mathrm{BML},i}$ is the melt fraction and $L_i$ is the specific latent heat.

Equations~\ref{equation8}–\ref{equation11} provide straightforward formulations for calculating the total amount of heat transferred from the core to the mantle. Because heat transfer across the core–mantle boundary is rapid at this stage, these values ultimately determine the melting fate of the solid lower mantle.

Based on the total required heat and the mantle’s final melting state, we implemented our Parameterized Mantle Melting Model. First, we partitioned the core’s heat content into calculation units along the core's adiabatic temperature profile, as illustrated by the red and yellow regions on the left side of Figure~\ref{fig:1}. For illustrative purposes, the gray lines in Figure~\ref{fig:1} correspond to cooling intervals of 100K. In our actual model calculations, however, each unit was defined by a $5\mathrm{K}$ decrement in temperature along the adiabatic gradient at the CMB. We further assume that heat is transferred in discrete steps, such that the heat in each unit must be fully released before the next unit begins transferring its heat. This assumption is reasonable when the step size is sufficiently small. The right panel of Figure~\ref{fig:1} shows the mantle melting state as excess heat from the superheated core enters the mantle. This additional heat drives the formation of a basal melting layer, which progressively cools and expands over time. The labeled stages (1, 2, and 3) in the right panel illustrate the sequential development of this melting layer, while the red arrows indicate the direction of its upward and outward evolution.


Therefore, the heat flux from the core to the mantle is a key factor controlling the mantle's melting rate and the overall melting process. To calculate this heat flux, we employed the formulation developed by \cite{ke2009coupled}:

\begin{equation}
F_c \approx a_c \cdot k \cdot \Delta T_c^{4/3} \left(\frac{\alpha \cdot \rho_m \cdot g}{\kappa \cdot \mu_{BML}}\right)^{1/3}
\label{equation12}
\end{equation}

where \( F_c \) is the heat flux from the core to the mantle, \( a_c \) is a constant related to the scaling of heat flux in the parameterized convection model, \( \Delta T_c \) is the temperature difference between the core and the mantle, \( \rho_m \) is the mantle density, \( g \) is the gravitational acceleration, \( k \) is the thermal conductivity, \( \kappa \) is the thermal diffusivity, \( \alpha \) is the thermal expansivity, and \( \mu_{\mathrm{BML}} \) is the viscosity of the convective interior of the basal melt layer (BML). The coefficient \( a_c \) is typically close to 0.1; here, we adopt \( a_c = 0.12 \). The driving temperature difference \( \Delta T_c \) depends on the convective regime and boundary conditions. At high temperatures, \( \Delta T_c \) at the \( i^\text{th} \) step can be approximated as:

\begin{equation}
\Delta T_{ci} \approx T_{CMBi} - T_{rheoi}
\label{equation13}
\end{equation}

\begin{equation}
T_{BMLi} = T_{CMBi} - 0.5 \times (T_{CMBi} - T_{rheoi})
\label{equation14}
\end{equation}

\begin{equation}
\mu_{BMLi} = \mu_0 \cdot e^{\left(\frac{A_v}{R_g \cdot T_{BMLi}}\right)}
\label{equation15}
\end{equation}

where \( T_{\mathrm{rheo}} \) is the critical temperature for the rheological transition, \( \mu_0 \) is the reference viscosity, \( A_v \) is the activation energy, \( R_g \) is the universal gas constant, and \( T_{\mathrm{BML}} \) is the temperature of the basal melt layer. The subscript \( i \) denotes the \( i^\text{th} \) step. Therefore, at the \( i^\text{th} \) step, the total heat content of the BML is given by:
\begin{equation}
Q_{\mathrm{tran}} = \int_0^{r_i} 4\pi r^2 \, \rho_{\mathrm{mantle}}(r) \cdot \left[ c_p \left( T_{\mathrm{cr}}(r) - T(r) \right) + 0.4 L \right] \, \mathrm{d}r
\label{equation16}
\end{equation}
where \( r_i \) is the radial position of the base of the BML at the \( i^\text{th} \) step,
\( \rho_{\mathrm{mantle}}(r) \) is the radial-dependent mantle density (in kg/m³),
\( c_p \) is the specific heat capacity (in J/kg·K),
\( T_{\mathrm{cr}}(r) \) and \( T(r) \) are the critical temperature and actual temperature profiles of the mantle, respectively,
and \( L \) is the latent heat of melting (in J/kg).

The formulas presented in Equations~\ref{equation12}--\ref{equation16} are no longer applicable when the temperature difference \( \Delta T_c \) is relatively small, as indicated by the yellow region in Figure~\ref{fig:1}. In such cases, the heat transfer can still be estimated using the approach described by~\cite{ke2009coupled}. All the parameters used in our parameterized mantle melting model are shown in Table~\ref{tab:symbols}.


\begin{table}[ht]
\centering
\caption{Symbol Definitions and Values}
\label{tab:symbols}
\scriptsize
\begin{tabular}{@{}lll@{}}
\toprule
\textbf{Symbol} & \textbf{Definition} & \textbf{Value \& Unit} \\
\midrule
$R_g$ & Universal gas constant & $8.3144$ J mol$^{-1}$ K$^{-1}$ \\
$Ra_{\text{crit}}$ & Critical Rayleigh number & 660 \\
$\beta$ & Exponent parameter & $\frac{1}{3}$ \\
$\alpha_m, \alpha_{\text{UM}}, \alpha_{\text{LM}}$ & Mantle thermal expansion coefficient & $3 \times 10^{-5}$ K$^{-1}$ \\
$\alpha_c$ & Core thermal expansion coefficient & $1 \times 10^{-5}$ K$^{-1}$ \\
$T_s$ & Surface reference temperature & 300 K \\
$g_s$ & Surface gravity acceleration & 9.8 m s$^{-2}$ \\
$g_{\text{UM}}$ & Upper mantle gravity acceleration & 9.8 m s$^{-2}$ \\
$g_{\text{LM}}$ & Lower mantle gravity acceleration & 10 m s$^{-2}$ \\
$g_c$ & Core gravity acceleration & 10 m s$^{-2}$ \\
$\kappa_m, \kappa_{\text{UM}}, \kappa_{\text{LM}}$ & Mantle thermal diffusivity & $1 \times 10^{-6}$ m$^{2}$ s$^{-1}$ \\
$k_{\text{UM}}$ & Upper mantle thermal conductivity & 4.2 W m$^{-1}$ K$^{-1}$ \\
$k_{\text{LM}}$ & Lower mantle thermal conductivity & 10.0 W m$^{-1}$ K$^{-1}$ \\
$\text{M}_e$ & Earth mass & $5.97 \times 10^{24}$ kg \\
$\text{M}_{\text{c}}$ & Core mass & $1.95 \times 10^{24}$ kg \\
$\text{M}_{\text{m}}$ & Mantle mass & $4.06 \times 10^{24}$ kg \\
$r_e$ & Earth's radius & 6371 km \\
$r_{\text{cmb}}$ & Core-mantle boundary radius & 3254 km \\
$r_{\text{ic\_e}}$ & Inner core radius & 1221 km \\
$\rho_m$ & Mantle average density & 4800 kg m$^{-3}$ \\
$\rho_c$ & Core center density & 11900 kg m$^{-3}$ \\
$c_{p_{\text{m}}}$ & Mantle specific heat capacity & 1265 J kg$^{-1}$ K$^{-1}$ \\
$c_{p_{\text{c}}}$ & Core specific heat capacity & 840 J kg$^{-1}$ K$^{-1}$ \\
$\text{v}_e$ & Reference viscosity & $2 \times 10^{17}$ m$^{2}$ s$^{-1}$ \\
$\text{A}_{\text{v}}$ & Activation energy & $3 \times 10^{5}$ J mol$^{-1}$ \\
$\text{v}_{\text{ref}}$ & Calculated reference viscosity & $5 \times 10^{7}$ \\
$k_b$ & Boltzmann constant & $1.38 \times 10^{-23}$ m$^{2}$ kg s$^{-2}$ K$^{-1}$ \\
$T_{\text{cr}}$ & Critical temperature & 2530 K \\
$\mu_{0l}$ & Liquid-phase reference viscosity & $1 \times 10^{-5}$ Pa s \\
$\mu_{0s}$ & Solid-phase reference viscosity & $4.27 \times 10^{10}$ Pa s \\
$E_l$ & Liquid-phase activation energy & $1.95 \times 10^{5}$ J mol$^{-1}$ \\
$V_l$ & Liquid-phase activation volume & $-3.65 \times 10^{-6}$ m$^{3}$ mol$^{-1}$ \\
$L_{\text{m}}$ & Mantle latent heat & $5.2 \times 10^{6}$ J kg$^{-1}$ \\
$R_c$ & Core radius & 3480 km \\
$D_N$ & Core adiabatic length scale & 6340 km \\
\bottomrule
\end{tabular}
\end{table}


\section{Results} \label{sec:3}
\subsection{Giant Impact-Induced Initial Mantle Melting} \label{sec:3.1}

Giant impact simulations have demonstrated that collisions can generate substantial heat, potentially causing extensive mantle melting \citep{nakajima2015melting,nakajima2021scaling}. In this study, we systematically investigate the initial mantle melting induced by collisions, building upon a set of 60 simulations of Earth-sized impacts and two models of the Moon-forming giant impact. These simulations span a wide range of impact angles (15\textdegree--75\textdegree), impact velocities (1--3~$V_{\text{esc}}$), and impactor masses (0.05--0.3~$M_{\oplus}$), thus covering a broad range of conceivable impact conditions. We employed the Gasoline Smoothed Particle Hydrodynamics (SPH) code, incorporating density corrections at silicate–metal boundaries and utilizing the latest equations of state \citep{stewart_sarah_t_2019,sarah_t_stewart_2020}. These enhancements enable more accurate estimates of temperature and pressure compared to standard SPH approaches \citep{ruiz2022dealing, zhou2024scaling}.

\begin{figure}[ht!]
\includegraphics[scale=1.1]{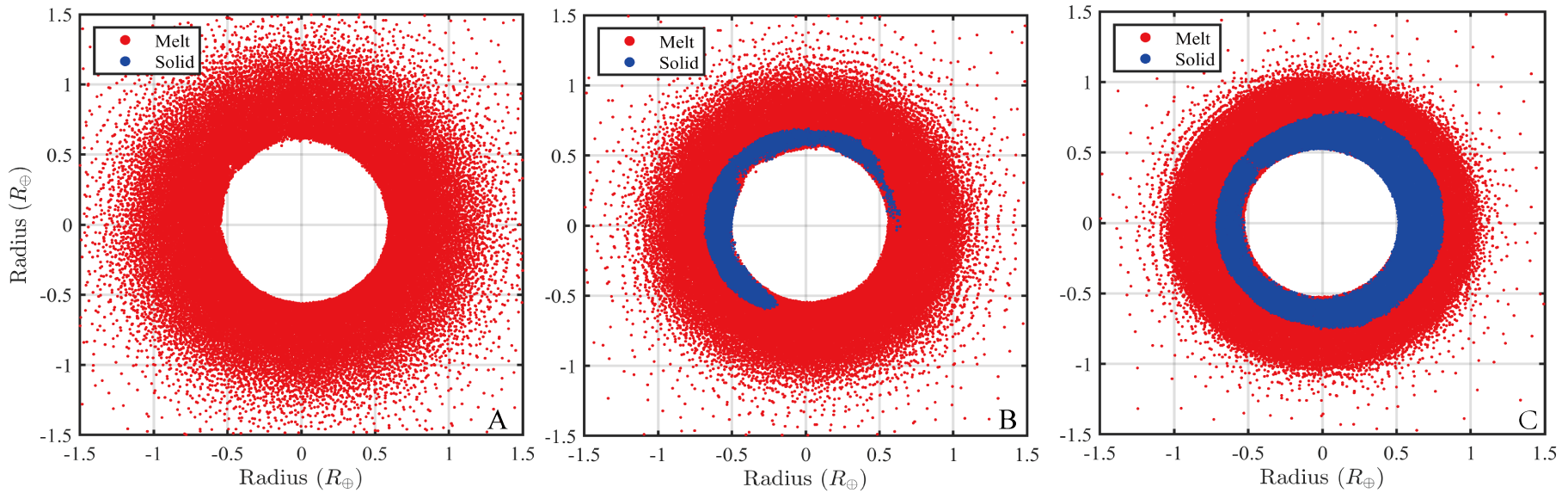}
\caption{\textbf{Typical Giant Impact-Induced Mantle Melting States.} This figure presents a cross-sectional view of three representative mantle melting states, as shown in Panels A, B, and C. The mantle particles are confined within a radial distance of $-0.1R_{\oplus} < R < 0.1R_{\oplus}$ along the Z-axis and projected onto the X–Y plane. Blue particles represent solid-phase material, while red particles indicate molten material or regions exceeding the melting temperature. The melting criterion is determined by Equations~\ref{equation17}–\ref{equation18}.
}
\label{fig:2}
\end{figure}

Figure \ref{fig:2} shows typical impact-induced mantle melting configurations after collisions. Panels A, B, and C illustrate three distinct post-impact mantle states: fully molten, partially molten, and upper-mantle molten, respectively. In these three scenarios, the target mass is fixed at 1$M_{\oplus}$, the impactor mass at 0.2$M_{\oplus}$, and the impact velocity at 2$V_{\text{esc}}$. The impact angles vary across the panels: 15° in the left panel, 30° in the middle panel, and 45° in the right panel.

We adopted the melting criterion from \cite{nakajima2021scaling}, defined as follows:

\begin{equation}
    \text{P} \leq 24 \text{ GPa}, \text{Tmelt}[K] = 1874 + 55.43\text{P} - 1.74\text{P}^2 + 0.0193\text{P}^3
    \label{equation17}
\end{equation}

\begin{equation}
    \text{P} > 24 \text{ GPa}, \text{Tmelt}[K] = 1249 + 58.28\text{P} - 0.395\text{P}^2 + 0.0011\text{P}^3
    \label{equation18}
\end{equation}

Here, P represents pressure and $\text{Tmelt}[K]$ represents the melting temperature.

We analyzed the mantle melting states across all simulations, as summarized in Appendix Table~\ref{tab:summary} and detailed in the Supplementary Materials. Here, we highlight two representative Moon-forming giant impacts: the Sub-Earth impact (Appendix Fig.~S1) and the canonical impact (Fig.~\ref{fig:4}). The Sub-Earth impact results in complete mantle melting, whereas the canonical impact leads to partial melting, characterized by a solid lower mantle and a molten upper mantle.

Based on all the initial mantle melting results, we find a clear dependence of mantle melting on the impact angle. When the angle exceeds $45^\circ$, the target planet generally retains most of its primordial lower mantle. In contrast, for impacts below $45^\circ$, preservation of the lower mantle becomes increasingly unlikely. At $45^\circ$, impacts typically produce partial melting restricted to the upper mantle, while the lower mantle remains largely solid. In addition to impact angle, impact velocity also plays a critical role in determining the extent of mantle melting. Higher velocities generally produce more extensive melting. However, in some cases, lower-velocity collisions (e.g., $2V_{\text{esc}}$) produce more melting than higher-velocity ones (e.g. $3V_{\text{esc}}$) because the latter often result in hit-and-run encounters that transfer less energy to the target. The impactor’s mass exerts a similar but less pronounced influence on mantle melting compared to impact velocity.

Given that impact velocities during Earth's late-stage accretion were typically only slightly higher than $1V_{\text{esc}}$, the most probable impact angle was approximately $45^\circ$ \citep{pierazzo2000melt}, and the most likely impactor-to-target mass ratio was in the range of 0.2–0.25 \citep{agnor1999}, our results imply that complete initial mantle melting was limited during Earth-sized giant impacts.

\subsection{Superheated Core-Induced Secondary Mantle Melting} \label{sec:3.2}

\begin{figure}[ht!]
\plotone{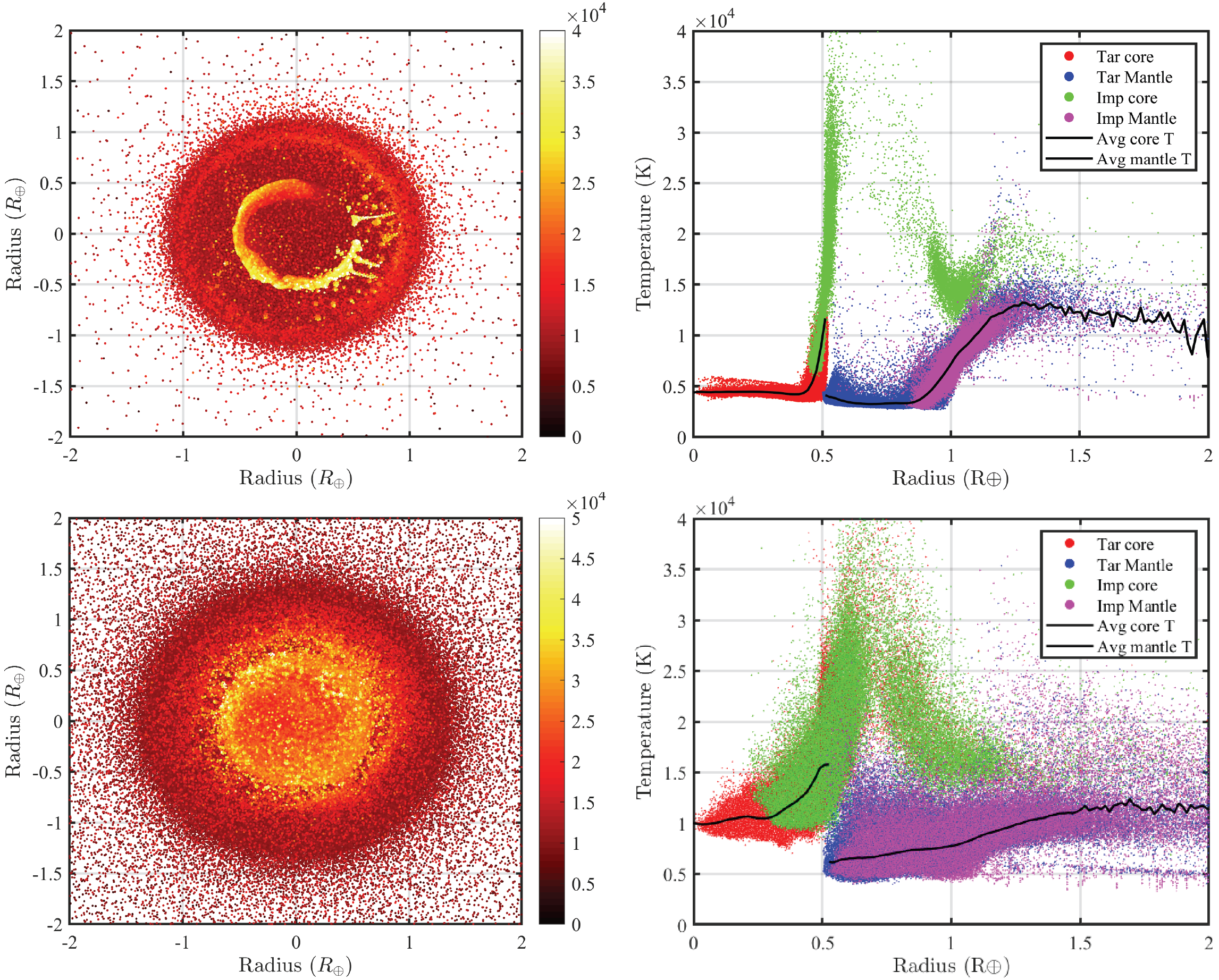}
\caption{\textbf{Thermal state of the proto-Earth following Moon-forming giant impacts.} The top row shows results from the canonical impact model, while the bottom row corresponds to the sub-Earth impact model. In each row, the left panel presents a two-dimensional temperature distribution, with the color bar indicating temperature. The right panel displays the radial profile of mean temperature, where particles are color-coded according to their material origin: target core, target mantle, impactor core, and impactor mantle. Black lines indicate the average temperature profiles of the core and mantle. Impact parameters are listed in Supplementary Table~\ref{tab:summary}, consistent with previous Moon-forming giant impact models~\citep{canup2001origin,canup2012forming}.
}
\label{fig:3}
\end{figure}

\citet{zhou2024scaling} showed that a giant impact can generate a superheated core—a high-temperature region forming in the outer portion of Earth’s core. This arises because Theia’s material, which carried the highest energy during the Earth–Theia collision, was significantly hotter than the proto-Earth’s material (see Supplementary Fig.~S2). As a result, the post-impact thermal state of the Earth is primarily determined by the distribution of Theia’s material. After core merging, Theia’s metallic core tends to settle in the outermost region of Earth's core. Meanwhile, shock heating from Theia’s core material decays exponentially with distance, thereby preferentially heating Earth’s core particles near the CMB. This process leads to the natural formation of a superheated outer core region, as illustrated in Figure~\ref{fig:3}. Through systematic simulations of Earth-sized giant impacts, \citet{zhou2024scaling} consistently observed high-temperature zones near the CMB (see Supplementary Fig.~S3), suggesting that a superheated core is a natural outcome of giant impacts.

Figure~\ref{fig:3} illustrates two representative Moon-forming giant impact scenarios: the canonical model (top row) and the sub-Earth model (bottom row). In both cases, Theia’s core material is primarily located in the upper region of the proto-Earth’s core, spatially coinciding with the superheated zones shown in the left panels. These simulation results are consistent with the superheating mechanism discussed above. In the canonical impact scenario, the temperature at the CMB reaches 11,663~K on the core side and 4,111~K on the mantle side. In the sub-Earth scenario, these values increase to 15,769~K and 6,122~K, respectively. Despite the differences in impact energy and heating intensity, both Moon-forming models robustly produce a superheated core and substantial temperature gradients across the CMB.

It is important to note that the superheated core would form a stable thermal stratification \citep{zhou2024scaling}, which significantly suppresses heat transport into the deeper core \citep{arkani2010giant}. As a result, excess heat from the super-heated core would be primarily transferred into the overlying mantle. This heat flux would initially melt the lowermost mantle, and the melting rate would exceed the growth rate of convective instabilities in the melt layer \citep{Tackley2025,ke2009coupled}. Consequently, a basal melt layer forms rapidly at the mantle base and may expand upward as melting proceeds. Two possible outcomes can arise from this process: (1) if the heat transferred from the core exceeds the energy required to melt the remnant mantle, the entire mantle would melt; (2) if the heat is insufficient, a melt layer would form at the base of the mantle. This basal melt layer could then evolve along two distinct pathways. One possibility is that instabilities within the basal melt layer would trigger the emergence of an early superplume or piles \citep{ke2009coupled,Tackley2025}. Alternatively, if the melt layer remains gravitationally and thermally stable, it could develop into a long-lived basal magma ocean that influences Earth’s long-term thermal and chemical evolution.

For all modeled impact scenarios, we used the Parameterized Mantle Melting Model developed in this study to quantify the heat released from core superheating, the energy required to melt the residual solid mantle, the duration of the melting phase, and the resulting mantle melting state (Table~\ref{tab:summary}). The temporal evolution of these melting processes is illustrated in Supplementary Fig.~S4. As an illustrative case, the secondary (post-impact) mantle melting following the canonical giant impact is analyzed in detail in Section~\ref{sec:3.3}.


\subsection{Earth's Magma Ocean After the Canonical collision} \label{sec:3.3}

Our simulations show that high-energy giant impacts (e.g., \citep{canup2012forming, cuk2012making}) can generate a globally molten magma ocean (see Supplementary Fig.~S1), whereas lower-energy impacts (e.g., \citep{canup2001origin}) melt only the upper mantle, leaving the lower mantle largely solid (see Fig.~\ref{fig:2} and Fig.~\ref{fig:3}). The presence of this residual solid lower mantle raises an important question: to what extent can core superheating remelt the lower mantle following the canonical collision? To answer this, we examine the secondary mantle-melting process and the subsequent evolution of the post-impact magma ocean in detail.

The secondary mantle melting driven by a superheated core is primarily governed by the viscosity of the basal melt layer, rather than that of the overlying solid mantle. A low-viscosity basal layer allows heat to be transferred efficiently from the core to the mantle. Parameterized convection models have estimated that this transfer can occur on timescales ranging from less than 100 years \citep{ke2009coupled} to about 1 million years \citep{andrault2016deep}, depending on assumptions about the physical state of the basal melt layer. Despite this broad range, both studies agree that heat exchange between the core and mantle proceeds rapidly.

In our simulations, the post-impact temperature at the core–mantle boundary (CMB) substantially exceeds the silicate liquidus, indicating that the lowermost mantle was likely partially or even fully molten. Under such conditions, the viscosity of the basal melt layer would decrease by several orders of magnitude, allowing highly efficient thermal coupling between the core and mantle. Consequently, the timescale for heat redistribution across the basal boundary layer in our model falls within the rapid convective regime inferred by \citet{ke2009coupled}.

To quantify this process, we first calculated the total heat released from the core and the heat required to melt the remnant lower mantle using Equations~\ref{equation8} through~\ref{equation11}. The results are summarized in Table~\ref{tab:summary}. Figure~\ref{fig:1} schematically illustrates the melting process: when the core temperature greatly exceeds that of the mantle, heat is transferred from the red region (core) to the green region (mantle), and the resulting CMB heat flux is computed using Equations~\ref{equation12} through~\ref{equation16}. If the heat content of the core (red region) is lower than that of the surrounding layer (blue region), the yellow region continues to transfer heat under conditions of small temperature gradients \citep{ke2009coupled}. Specifically, for the canonical impact scenario, the superheated core must transfer approximately $1.3152\times10^{30}$~J of heat to the mantle, whereas $3.4021\times10^{30}$~J is required to raise the lower mantle from the solidus to the rheological transition temperature. This result implies that the heat released from the core alone is insufficient to completely melt the lower mantle following the canonical Moon-forming impact (see Table~\ref{tab:summary}).

\begin{figure}[ht!]
\plotone{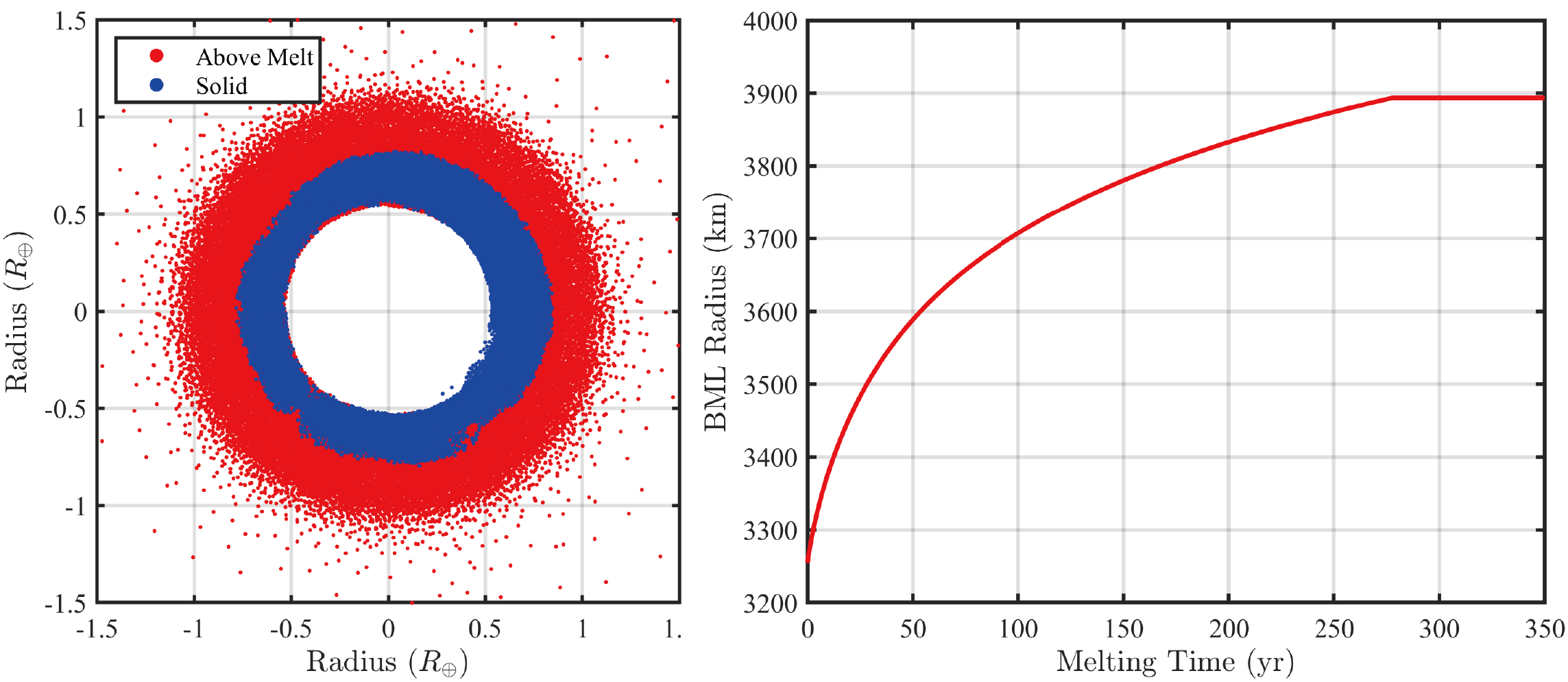}
\caption{ \textbf{Melting of the solid lower mantle induced by a superheated core following the canonical impact.} The left panel shows the post-impact mantle state, where the lower mantle remains solid. Simulation parameters are listed in Table \ref{tab:symbols}. The right panel illustrates the melting of the lower mantle driven by heat transfer from the superheated core. The horizontal axis denotes the melting timescale, and the vertical axis (BML radius) represents the distance from the core to the upper boundary of the basal melt layer (BML).}
\label{fig:4}
\end{figure}

We further employed our parameterized mantle melting model to simulate the post-impact evolution of the mantle following the canonical giant collision. The model was adapted to match the specific initial and boundary conditions described in the Methods section. Figure~\ref{fig:4} presents the results. The left panel shows the immediate post-impact state, where the proto-Earth’s upper mantle is entirely molten while the lower mantle remains solid. At this stage, the thickness of the solid lower mantle defines the upper and lower boundaries of the basal melt layer used in subsequent calculations (see Supplementary Fig.~S5). The right panel illustrates the progressive bottom-up melting of the lower mantle driven by heat transfer from the superheated core. Our results indicate that portions of the lower mantle can melt within approximately 277~years, forming a basal melt layer about 350~km thick (Fig.~\ref{fig:4}). This timescale lies well within the rapid convective regime predicted by parameterized convection models, suggesting that the proto-Earth likely experienced partial remelting of the lower mantle following the canonical Moon-forming impact.

\begin{figure}[ht!]
\centering
\includegraphics[width=0.7\textwidth]{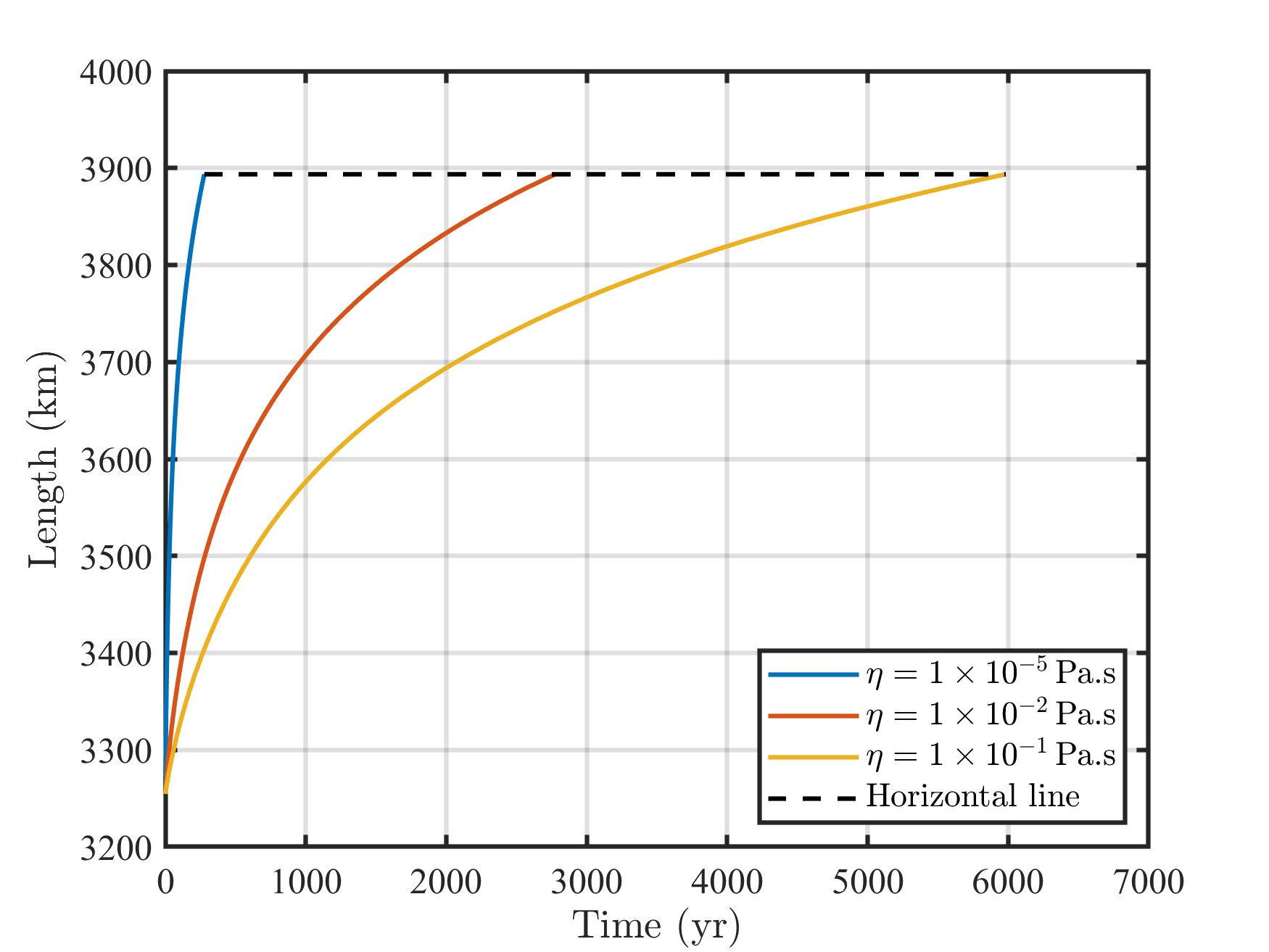}
\caption{\textbf{Melting process of the lower mantle under the different viscosity assumption.} Different colored lines represent different viscosities, resulting in varying melting times and melting behaviors.}
\label{fig:5}
\end{figure}

The melting rate of the lower mantle is strongly influenced by the viscosity of the partially molten region, which remains highly uncertain. To assess this effect, we tested several viscosity values, including $1\times10^{-5}$~Pa~s~\citep{ke2009coupled}, $1\times10^{-2}$~Pa~s~\citep{monteux2016}, and $1\times10^{-1}$~Pa~s~\citep{dingwell2004viscosity}. The corresponding melting processes are shown in Figure~\ref{fig:5}. As illustrated, the duration of melting ranges from approximately 277~years to 5983~years.

It should be noted that these results depend sensitively on the impact conditions and thus carry substantial uncertainties (see Section~\ref{sec:4.3}). Variations in these parameters could lead to more extensive mantle melting, potentially generating a fully molten magma ocean. Despite these uncertainties, our findings provide a testable framework: present-day geophysical and geochemical observations of the deep Earth may offer critical constraints on whether such a layer may have once existed during Earth’s early evolution.

\section{Disscussion} \label{sec:4}
\subsection{Short-Term Impact: Disrupting Primordial Lower Mantle Heterogeneity} \label{sec:4.1}

Recent simulations of the Moon-forming giant impact suggest that the canonical scenario primarily melted Earth’s upper mantle, whereas much of the lower mantle remained largely solid \citep{nakajima2021scaling,deng2019primordial,lock2020geochemical}. Portions of this unmelted lower mantle may have survived subsequent mantle convection and mixing, thereby resisting complete homogenization over geologic time \citep{ballmer2017persistence,pachhai2022internal}. These dynamical predictions are consistent with extensive isotopic evidence for long-lived mantle heterogeneity. Such heterogeneity was first recognized in the early 1960s through studies of mid-ocean ridge basalts (MORBs) and ocean island basalts (OIBs), which indicated the existence of at least two compositionally distinct mantle reservoirs that have remained only partially mixed for over 4.5 billion years (e.g., \citep{caro2003146sm,mukhopadhyay2012early,touboul2015tungsten,willbold2011tungsten,willbold2015tungsten,walker1995osmium,rizo2019182w,mundl2020anomalous}). In particular, many OIBs sourced from deep mantle plumes carry isotopic signatures that point to very ancient reservoirs still preserved in the lower mantle. Some of these primordial materials may have formed as early as 4.45–4.55 billion years ago, coincident with the Moon-forming time \citep{jackson2010evidence,horton2023highest}.

This consensus between dynamical simulations and isotopic evidence strongly suggests that portions of Earth’s lower mantle survived the Moon-forming collision and have persisted to the present day. Building on this insight, \cite{yuan2023} proposed that remnants of Theia might have been preserved in the lower mantle after the Moon-forming collision, subsequently becoming the main components of Earth’s Large Low Shear Velocity Provinces (LLSVPs). In this view, the primordial heterogeneity of the lower mantle provides a key constraint on Moon-forming giant impact scenarios \citep{halliday2023accretion}. By contrast, alternative high-energy impact models would likely have generated a fully molten magma ocean (e.g., \citep{canup2012forming,cuk2012making}), erasing any pre-existing chemical reservoirs in the deep mantle (see Fig. S1). It should be noted, however, that some studies argue that such isotopic anomalies may not solely reflect ancient mantle differentiation. Instead, they could also arise from long-term core–mantle chemical exchange, which can generate “primordial-like” isotopic signatures even in the absence of truly ancient mantle domains \citep{porcelli2001core,ferrick2023long}.

However, the superheated core would have naturally destroyed primordial heterogeneities in the lower mantle, as heat transfer from the core would drive bottom-up melting and generate a vigorously convecting basal melt layer. In this region, any pre-existing reservoirs would be homogenized. If the basal melt layer did not completely melt the solid lower mantle, some heterogeneities might have temporarily survived within the mid-mantle. Nevertheless, such reservoirs would likely have been unstable over the long term, as mid-mantle convection would readily disrupt them, and primordial heterogeneities could only be preserved near the mantle’s base \citep{yuan2023}.

Figure~\ref{fig:5} presents a schematic illustration of mantle melting driven by a superheated core after a giant impact. Panel~A shows the initial post-impact thermal structure of Earth’s mantle following a canonical collision, with a molten upper mantle overlying a solid lower mantle that retains some primordial heterogeneous reservoirs, while the core remains at very high temperature (specific values shown in Figure~\ref{fig:3}). As the system evolves (Panel~B), the upper molten layer cools toward the rheological transition temperature \citep{monteux2016}, but the large temperature contrast between the hot core and the mantle triggers bottom-up melting, generating a basal melt layer at the mantle’s base in which most primordial heterogeneities are likely destroyed. In the final stage (Panel~C), as the mantle continues to cool, part of the basal melt may crystallize or be lost, yet at least a small-scale basal magma ocean is expected to persist while the rest of the mantle solidifies.

In summary, the secondary melting induced by the superheated core would largely destroy the primordial heterogeneity of the lower mantle, implying that the oldest reservoirs preserved in the present-day mantle must have originated after the Moon-forming event.

\begin{figure}[ht!]
\plotone{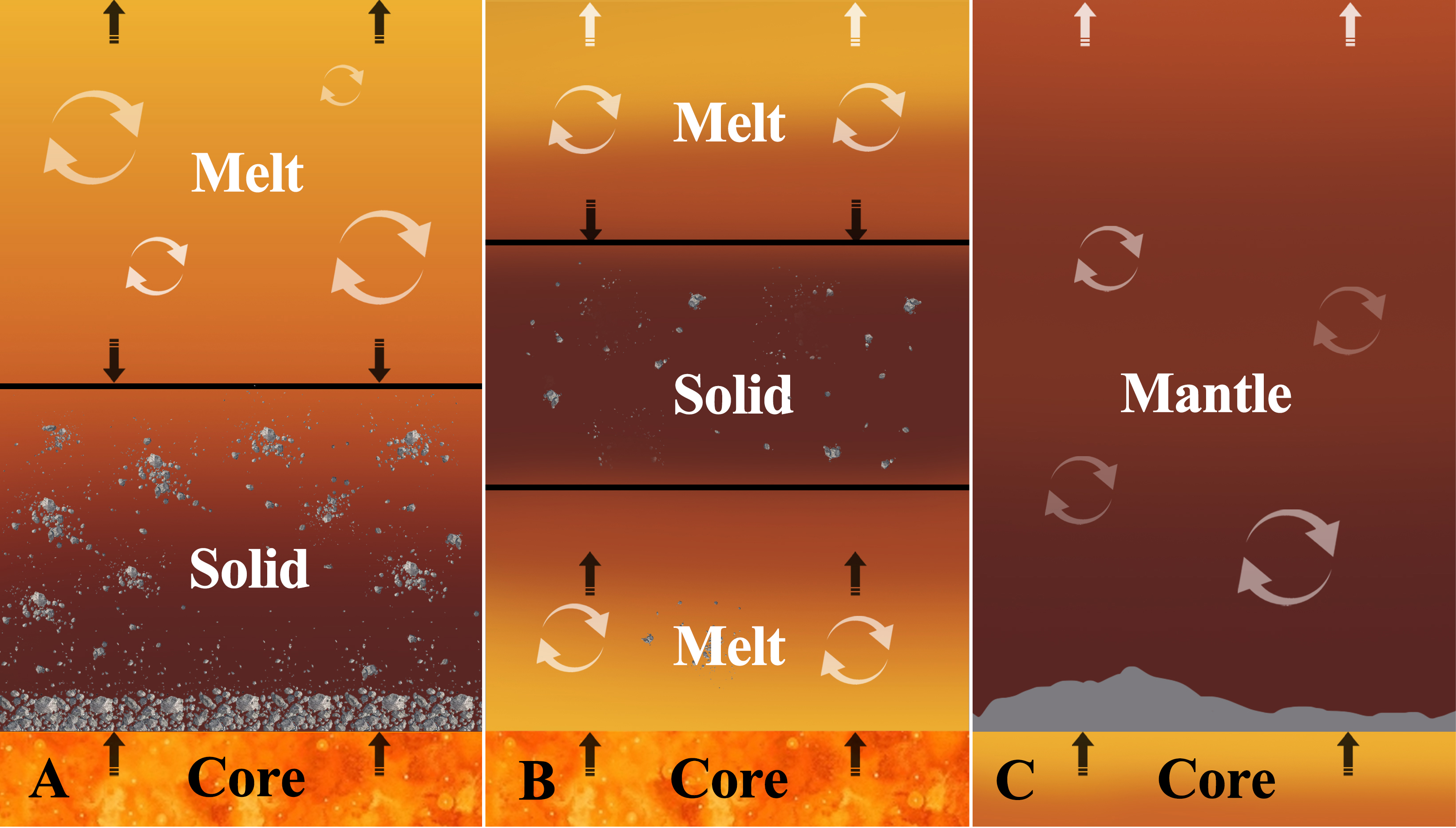}
\caption{\textbf{A schematic diagram of the superheated core melting Earth's entire mantle.}
The figure illustrates the progressive melting of Earth’s mantle driven by heat transfer from a superheated core following a canonical giant impact. “Melt” denotes the molten portion of the mantle, “Solid” the unmelted mantle, and “Core” the metallic core. The silvery materials within the mantle represent primordial heterogeneities in the lower mantle. Panels~A–C depict the temporal evolution of the mantle’s thermal structure. This schematic highlights the link between core–mantle heat exchange, large-scale mantle melting, and the potential destruction of early-formed heterogeneities.}
\label{fig:6}
\end{figure}

\subsection{Long-Term Impact: Formation of a Basal Melt Layer} \label{sec:4.2}

\citet{ke2009coupled} and \citet{Tackley2025} proposed that a superheated core could lead to rapid bottom-up melting of the solid mantle, generating a rapidly expanding mantle melt layer and potentially triggering an early-stage superplume event. \citet{nakagawa2010influence} further showed that, for an Earth-like planet, models with different initial core temperatures tend to converge toward a similar thermal evolutionary path. Therefore, \citet{Tackley2025} suggested that the effect of core superheating on Earth's thermal evolution is likely to be short-lived. Here, we discuss this issue in more detail.

Fundamentally, the long-term influence of core superheating depends on the solidification process of Earth's magma ocean, which remains an unresolved problem due to the high Rayleigh number associated with such systems. However, in recent years, significant progress has been made in modeling two-phase mantle flow. \citet{boukare2017modeling} developed a two-phase numerical model and argued that the density contrast between solid and melt is as critical as the intersection between the liquidus and the isentrope. \citet{boukare2025solidification} further showed that a basal magma ocean (BMO) can inevitably form due to the gravitational segregation of iron-rich melts, regardless of where the melting curve intersects the geotherm. These recent two-phase flow studies not only provide a self-consistent framework for understanding magma ocean solidification, but also establish a basis for inferring the final state of the basal molten layer.

Based on the secondary melting results in Table~\ref{tab:summary}, we find that the basal molten layer may evolve into three distinct scenarios: (1) a fully molten magma ocean; (2) an early superplume event; or (3) a stable basal magma ocean (BMO). By linking the initial post-impact mantle melting state with recent advances in two-phase flow modeling, we can infer the thermal evolution of the basal molten layer under these three conditions, as illustrated in Figure~\ref{fig:6}. In Figure~\ref{fig:6}, \ding{172} represents a scenario in which core superheating triggers complete mantle melting, leaving only a small basal magma ocean after subsequent cooling \citep{boukare2025solidification}. \ding{173} corresponds to a case in which the basal molten layer is thermally unstable and rapidly releases its energy through an early superplume or mantle piles, leaving a partially preserved melt layer \citep{boukare2017modeling,ke2009coupled}. Finally, \ding{174} depicts a scenario in which the basal molten layer remains thermally stable and develops into a widespread, long-lived basal magma ocean.

\begin{figure}[ht!]
\includegraphics[scale=0.6]{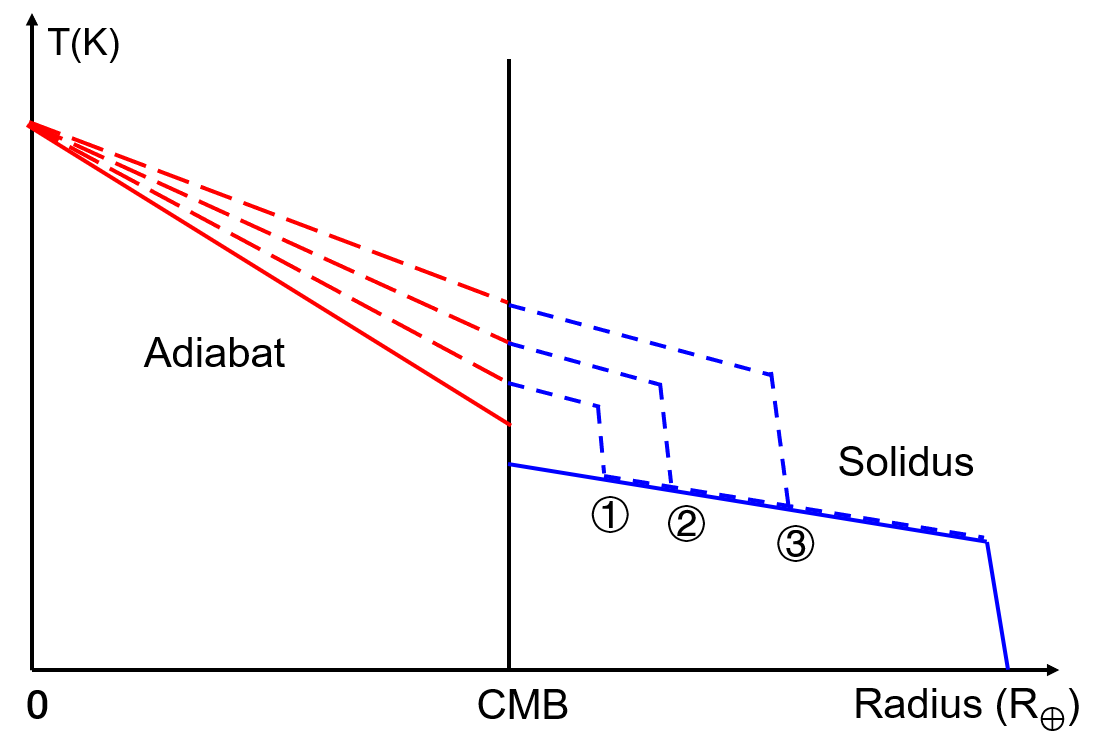}
\centering
\caption{\textbf{Schematic illustration of the possible evolutionary outcomes of the basal molten layer.}
Labels \ding{172}–\ding{174} correspond to the three thermal evolution scenarios discussed in the text. The red solid line represents the mantle's adiabatic temperature profile, while the three red dashed lines indicate potential core temperature paths associated with each scenario. The blue solid line denotes the mantle solidus, and the blue dashed line shows possible mantle temperature profiles. This figure is for illustrative purposes and does not represent simulation results.}
\label{fig:7}
\end{figure}

In summary, our analysis indicates that the solid lower mantle would have melted rapidly after the canonical collision, making the formation of a Basal Magma Ocean (BMO) highly probable. Such a BMO would not only serve as a heat reservoir but, more importantly, would elevate the core temperature profile and inhibit the loss of core superheating (Fig.~\ref{fig:6}), thereby shaping Earth’s long-term thermal evolution.

\cite{zhou2024scaling} suggest that a superheated core could delay the onset of the geodynamo until the core cooled to an adiabatic state. Combined with the earliest geomagnetic evidence preserved in zircon inclusions dated to ~4.2 Ga~\citep{tarduno2020}, this provides a potential constraint on the Moon-forming giant impact. The interval between this impact and the oldest zircon-recorded magnetic field is only ~250 Myr, implying that any basal magma ocean (BMO) present must have been relatively small in volume and short-lived. This interpretation is consistent with the view that Earth initially hosted a fully molten magma ocean that cooled to form only a thin BMO (Scenario \ding{172}, Fig.~\ref{fig:6}). However, the timing of the earliest geodynamo remains debated \citep{borlina2020reevaluating}. If the dynamo instead initiated at ~3.5 Ga \citep{tarduno2010geodynamo, biggin2015palaeomagnetic}, then core superheating could have sustained a medium-sized BMO in the lower mantle, which later lost substantial heat through a superplume or mantle piles (Scenario \ding{173}, Fig.~\ref{fig:6}).

\subsection{Uncertainty in Modeling Core Superheating} \label{sec:4.3}

This study systematically investigates the thermal state resulting from Earth-sized collisions, providing a foundation for future studies on planetary thermal evolution. In particular, for the canonical impact scenario, our results suggest that core superheating leads to partial melting of the residual mantle within 277--5983~years, though complete melting of the mantle is also a possibility. However, this conclusion remains sensitive to several other factors.

First, the initial thermal states of proto-Earth and Theia remain uncertain. In this study, we assume an initial surface temperature of 2000~K for the proto-Earth and 1000~K for Theia, with internal thermal profiles constructed along adiabats anchored at these surface temperatures. In reality, the initial thermal states of both bodies are poorly constrained, and variations in their initial temperatures could significantly alter the degree of mantle melting following the giant impact.

Second, the canonical impact scenario spans a range of plausible parameters. For example, the mass ratio between Theia and the proto-Earth may vary from 0.13 to 0.2, and the impact velocity can range from 1.0 to 1.2 times the mutual escape velocity ($V_{\text{esc}}$). While these values are generally accepted within the context of lunar formation, they can lead to considerable differences in the resulting thermal state of the post-impact Earth.

Third, the mantle rheological transition temperature adopted in this study was taken from \citep{monteux2016}. However, there exist a range of alternative estimates: higher values reported by \citep{miyazaki2019timescale} and lower values reported by \citep{stixrude2015seismic}. The chosen rheological transition temperature directly determines the amount of core superheating that the mantle can accommodate. A related key parameter is the mantle latent heat. In this study, we adopted a value of $5.2 \times 10^{6}$~J/kg from \citep{rubie2015accretion}, while other estimates include a lower value of $3.2\times 10^{5}$~kJ/kg \citep{driscoll2014thermal} and a higher value of $9 \times 10^{6}$~J/kg \citep{korenaga2023rapid}.

Finally, the standard Smoothed Particle Hydrodynamics (SSPH) method introduces numerical artifacts at the CMB~\citep{ruiz2022dealing,zhou2024scaling}. Because temperature is highly sensitive to spatial resolution near such discontinuities, these numerical errors can lead to an underestimation of the temperature at the CMB~\citep{zhou2021core}, potentially affecting the evaluation of whether the mantle becomes fully molten.

\section{Conclusions} \label{sec:5}

In this study, based on detailed SPH simulations of Earth-sized impacts and a parameterized mantle melting model, we propose that core superheating plays a decisive role in the subsequent evolution of Earth's mantle. It may lead to one of three outcomes: the formation of a fully molten magma ocean, the triggering of an early superplume, or the formation of a basal melt layer. Therefore, the superheated core, in the short term, may destroy any primordial heterogeneities and lead to a chemically more homogeneous mantle. These dynamic results suggest that the formation of present-day mantle heterogeneities is unlikely to precede the Moon-forming impact event.

Beyond the immediate post-impact effects, core superheating may also influence Earth's long-term thermal evolution via the formation of a basal magma ocean. Although this basal melt layer may dissipate rapidly through superplume activity \citep{ke2009coupled, Tackley2025}, the latest two-phase flow simulations suggest that the formation of a basal magma ocean is inevitable \citep{boukare2025solidification}. This implies that the final evolution of this basal layer would inhibit core cooling and delay the onset of Earth's geodynamo. Given that superheated cores are likely a common consequence of giant impacts~\citep{marchi2023long, zhou2024scaling}, similar processes may have contributed to the formation of basal magma oceans on other terrestrial bodies, including Mars and the Moon.

Despite uncertainties—including initial thermal conditions, impact parameters, and numerical artifacts—our study highlights core superheating as a fundamental driver of both the short- and long-term thermal evolution of Earth. Future work should aim to couple giant impact simulations with two-phase mantle convection models to develop a more comprehensive understanding of the thermal and chemical resetting of the early Earth.

\subsection*{Acknowledgment}

We sincerely thank Dr.~Peter~Driscoll and Prof.~Yue~Zongyu for their constructive and insightful suggestions.

\section*{Data Availability Statement}

Some key computational results used in this study—including those from the primary and secondary melting schemes and the parameterized core-cooling model—are accessible at https://osf.io/za7hm/

\section*{Appendix}  
\setcounter{figure}{0}  
\renewcommand{\thefigure}{S\arabic{figure}}  

\subsection*{Appendix A: Simulation parameters and results}

\begin{table}[ht]
\centering
\caption{Simulation parameters and results summary}
\label{tab:summary}
\scriptsize
\resizebox{\textwidth}{!}{ 
\begin{tabular}{@{}ccccccccc@{}}
\toprule
\textbf{Case} & $\sin(\theta)$ & $\frac{M_i}{M_{\oplus}}$ & $\frac{V_i}{V_{\text{esc}}}$ & \textbf{Core Heating (J)} & \textbf{Mantle Melting Heat (J)} & \textbf{Mantle Melting Time (yr)}& \textbf{Initial Melting} & \textbf{Secondary Melting} \\
\midrule
Case1 & 0.2588 & 0.05 & 1 & $1.5173\times10^{30}$ & $1.9556\times10^{30}$ & $3.8856\times10^{2}$ & Partial Melting & Partial Melting \\
Case2 & 0.5000 & 0.05 & 1 & $1.2822\times10^{30}$ & $4.0380\times10^{30}$ & $2.9800\times10^{2}$ & Partial Melting & Partial Melting \\
Case3 & 0.7071 & 0.05 & 1 & $5.6134\times10^{29}$ & $5.1530\times10^{30}$ & $1.7060\times10^{2}$ & Partial Melting & Partial Melting \\
Case4 & 0.8660 & 0.05 & 1 & $3.5931\times10^{28}$ & $6.6245\times10^{30}$ & $1.6551\times10^{1}$ & Partial Melting & Partial Melting \\
Case5 & 0.9659 & 0.05 & 1 & \textemdash & $8.3414\times10^{30}$ & 0 & Partial Melting & Partial Melting \\
Case6 & 0.2588 & 0.05 & 2 & $5.0388\times10^{30}$ & 0 & 0 & Complete Melting & \textemdash \\
Case7 & 0.5000 & 0.05 & 2 & $6.6834\times10^{29}$ & $1.7646\times10^{30}$ & $2.4752\times10^{2}$ & Partial Melting & Partial Melting \\
Case8 & 0.7071 & 0.05 & 2 & $7.3800\times10^{26}$ & $4.5825\times10^{30}$ & $1.4474\times10^{1}$ & Partial Melting & Partial Melting \\
Case9 & 0.8660 & 0.05 & 2 & \textemdash & $7.8552\times10^{30}$ & 0 & Partial Melting & Partial Melting \\
Case10 & 0.9659 & 0.05 & 2 & \textemdash & $1.0131\times10^{31}$ & 0 & Partial Melting & Partial Melting \\
Case11 & 0.2588 & 0.05 & 3 & $9.0686\times10^{30}$ & 0 & 0 &  Complete Melting & \textemdash \\
Case12 & 0.5000 & 0.05 & 3 & $1.2073\times10^{30}$ & $1.2071\times10^{30}$ & $7.7020\times10^{2}$ & Partial Melting & Complete Melting \\
Case13 & 0.7071 & 0.05 & 3 & $1.3020\times10^{29}$ & $3.6125\times10^{30}$ & $5.6799\times10^{1}$ & Partial Melting & Partial Melting \\
Case14 & 0.8660 & 0.05 & 3 & \textemdash & $7.4815\times10^{30}$ & 0 & Partial Melting & Partial Melting \\
Case15 & 0.9659 & 0.05 & 3 & \textemdash & $1.0413\times10^{31}$ & 0 & Partial Melting & Partial Melting \\
Case16 & 0.2588 & 0.10 & 1 & $5.5559\times10^{30}$ & 0 & 0 & Complete Melting & \textemdash \\
Case17 & 0.5000 & 0.10 & 1 & $3.3137\times10^{30}$ & $1.6967\times10^{30}$ &$5.9347\times10^{2}$ & Partial Melting & Complete Melting \\
Case18 & 0.7071 & 0.10 & 1 & $1.2559\times10^{30}$ & $3.4594\times10^{30}$ & $2.8769\times10^{2}$ & Partial Melting & Partial Melting \\
Case19 & 0.8660 & 0.10 & 1 & $1.5785\times10^{29}$ & $5.5937\times10^{30}$ & $6.1200\times10^{1}$ & Partial Melting & Partial Melting \\
Case20 & 0.9659 & 0.10 & 1 & \textemdash & $7.6153\times10^{30}$ & 0 & Partial Melting & Partial Melting \\
Case21 & 0.2588 & 0.10 & 2 & $8.7816\times10^{30}$ & 0 & 0 & Complete Melting & \textemdash \\
Case22 & 0.5000 & 0.10 & 2 & $2.4319\times10^{30}$ & 0 & $ 0 $ & Complete Melting & \textemdash \\
Case23 & 0.7071 & 0.10 & 2 & $3.6719\times10^{28}$ & $3.4083\times10^{30}$ & $1.8518\times10^{1}$ & Partial Melting & Partial Melting \\
Case24 & 0.8660 & 0.10 & 2 & \textemdash & $7.1753\times10^{30}$ & 0 & Partial Melting & Partial Melting \\
Case25 & 0.9659 & 0.10 & 2 & \textemdash & $9.7817\times10^{30}$ & 0 & Partial Melting & Partial Melting \\
Case26 & 0.2588 & 0.10 & 3 & $1.2255\times10^{31}$ & 0 & 0 & Complete Melting & \textemdash \\
Case27 & 0.5000 & 0.10 & 3 & $5.9075\times10^{30}$ & 0 & 0 & Complete Melting & \textemdash \\
Case28 & 0.7071 & 0.10 & 3 & $7.0022\times10^{29}$ & $2.5846\times10^{30}$ & $2.1774\times10^{2}$ & Partial Melting & Partial Melting \\
Case29 & 0.8660 & 0.10 & 3 & \textemdash & $6.7023\times10^{30}$ & 0 & Partial Melting & Partial Melting \\
Case30 & 0.9659 & 0.10 & 3 & \textemdash & $1.0090\times10^{31}$ & 0 & Partial Melting & Partial Melting \\
Case31 & 0.2588 & 0.20 & 1 & $1.3865\times10^{31}$ & 0 & 0 & Complete Melting & \textemdash \\
Case32 & 0.5000 & 0.20 & 1 & $7.3559\times10^{30}$ & 0 & 0 & Complete Melting & \textemdash \\
Case33 & 0.7071 & 0.20 & 1 & $4.4044\times10^{30}$ & 0 & 0 & Complete Melting & \textemdash \\
Case34 & 0.8660 & 0.20 & 1 & $6.1623\times10^{27}$ & $4.6133\times10^{30}$ & $3.5086\times10^{0}$ & Partial Melting & Partial Melting \\
Case35 & 0.9659 & 0.20 & 1 & \textemdash & $6.4553\times10^{30}$ & 0 & Partial Melting & Partial Melting \\
Case36 & 0.2588 & 0.20 & 2 & $1.7455\times10^{31}$ & 0 & 0 & Complete Melting & \textemdash \\
Case37 & 0.5000 & 0.20 & 2 & $3.4614\times10^{30}$ & 0 & 0 & Complete Melting & \textemdash \\
Case38 & 0.7071 & 0.20 & 2 & $3.3908\times10^{29}$ & $2.3376\times10^{30}$ & $1.2091\times10^{2}$ & Partial Melting & Partial Melting \\
Case39 & 0.8660 & 0.20 & 2 & \textemdash & $6.2561\times10^{30}$ & 0 & Partial Melting & Partial Melting \\
Case40 & 0.9659 & 0.20 & 2 & \textemdash & $9.3180\times10^{30}$ & 0 & Partial Melting & Partial Melting \\
Case41 & 0.2588 & 0.20 & 3 & $1.5147\times10^{31}$ & 0 & 0 & Complete Melting & \textemdash \\
Case42 & 0.5000 & 0.20 & 3 & $6.8640\times10^{30}$ & 0 & 0 & Complete Melting & \textemdash \\
Case43 & 0.7071 & 0.20 & 3 & $7.7636\times10^{29}$ & $1.8589\times10^{30}$ & $3.3461\times10^{2}$ & Partial Melting & Partial Melting \\
Case44 & 0.8660 & 0.20 & 3 & \textemdash & $5.9375\times10^{30}$ & 0 & Partial Melting & Partial Melting \\
Case45 & 0.9659 & 0.20 & 3 & \textemdash & $9.7106\times10^{30}$ & 0 & Partial Melting & Partial Melting \\
Case46 & 0.2588 & 0.30 & 1 & $1.8860\times10^{31}$ & 0 & 0 & Complete Melting & \textemdash \\
Case47 & 0.5000 & 0.30 & 1 & $9.5985\times10^{30}$ & 0 & 0 & Complete Melting & \textemdash \\
Case48 & 0.7071 & 0.30 & 1 & $4.6519\times10^{30}$ & 0 & 0 & Complete Melting & \textemdash \\
Case49 & 0.8660 & 0.30 & 1 & $4.6588\times10^{28}$ & $3.9149\times10^{30}$ & $2.4079\times10^{1}$ & Partial Melting & Partial Melting \\
Case50 & 0.9659 & 0.30 & 1 & $9.6301\times10^{26}$ & $5.3380\times10^{30}$ & $5.7632\times10^{-1}$ & Partial Melting & Partial Melting \\
Case51 & 0.2588 & 0.30 & 2 & $1.8860\times10^{31}$ & 0 & 0 & Complete Melting & \textemdash \\
Case52 & 0.5000 & 0.30 & 2 & $4.9323\times10^{30}$ & 0 & 0 & Complete Melting & \textemdash \\
Case53 & 0.7071 & 0.30 & 2 & $3.2746\times10^{29}$ & $1.7127\times10^{30}$ & $1.3885\times10^{2}$ & Partial Melting & Partial Melting \\
Case54 & 0.8660 & 0.30 & 2 & \textemdash & $5.8411\times10^{30}$ & 0 & Partial Melting & Partial Melting \\
Case55 & 0.9659 & 0.30 & 2 & \textemdash & $8.7605\times10^{30}$ & 0 & Partial Melting & Partial Melting \\
Case56 & 0.2588 & 0.30 & 3 & $9.9177\times10^{29}$ & $1.4881\times10^{30}$ & $4.0335\times10^{2}$ & Partial Melting & Partial Melting \\
Case57 & 0.5000 & 0.30 & 3 & $8.2355\times10^{30}$ & 0 & 0 & Complete Melting & \textemdash \\
Case58 & 0.7071 & 0.30 & 3 & $1.2377\times10^{30}$ & $1.6252\times10^{30}$ & $4.6020\times10^{2}$ & Partial Melting & Partial Melting \\
Case59 & 0.8660 & 0.30 & 3 & \textemdash & $5.3633\times10^{30}$ & 0 & Partial Melting & Partial Melting \\
Case60 & 0.9659 & 0.30 & 3 & \textemdash & $9.2829\times10^{30}$ & 0 & Partial Melting & Partial Melting \\
Canonical & 0.7071 & 0.112 & 1 & $1.3152\times10^{30}$ & $3.4021\times10^{30}$ & $2.7770\times10^{2}$ & Partial Melting & Partial Melting \\
Sub-Earth & 0.5500 & 0.495 & 1.1 & $8.0987\times10^{30}$ & 0  & 0 & Complete Melting & \textemdash \\
\bottomrule
\end{tabular}
}
\end{table}

%

\vspace{5mm}




\newpage

\bibliography{ref}
\bibliographystyle{aasjournal}



\section{Supplementary}

\begin{figure}[ht!]
\centering
\includegraphics[scale=1]{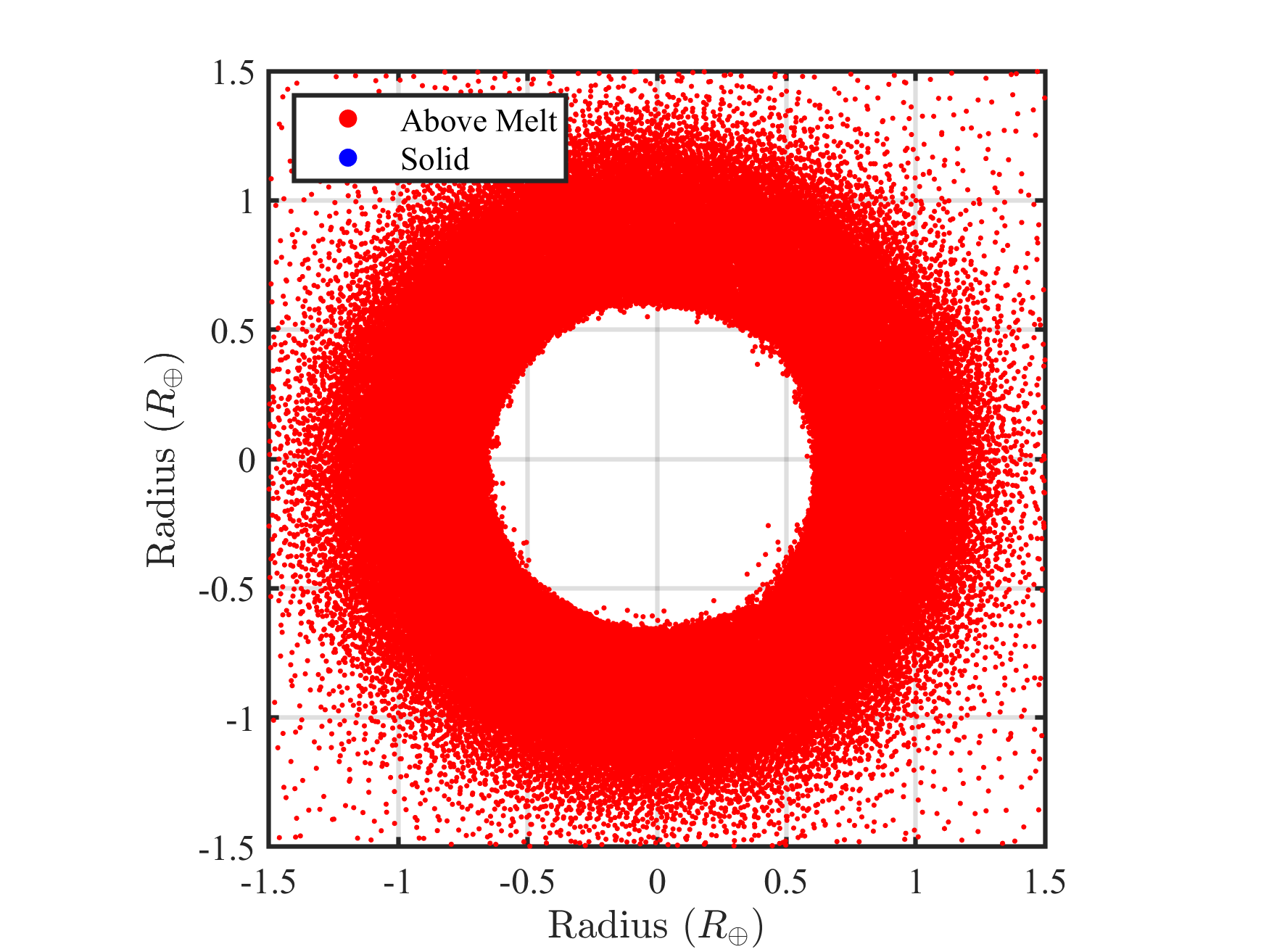}
\caption{The mantle melting state of the Sub-Earth collision. The Sub-Earth collision model was proposed by Canup(2012), with the initial collision conditions shown in Table~1. In the figure, the red particles represent melt and higher-temperature phases. All particles were sliced through the range $-0.1R_{\oplus} < R < 0.1R_{\oplus}$, and then projected onto the X-Y plane. The criterion for melting is based on Equation~(1).
\label{fig:s1}}
\end{figure}

\begin{figure}[!ht]
\centering
\includegraphics[scale=1]{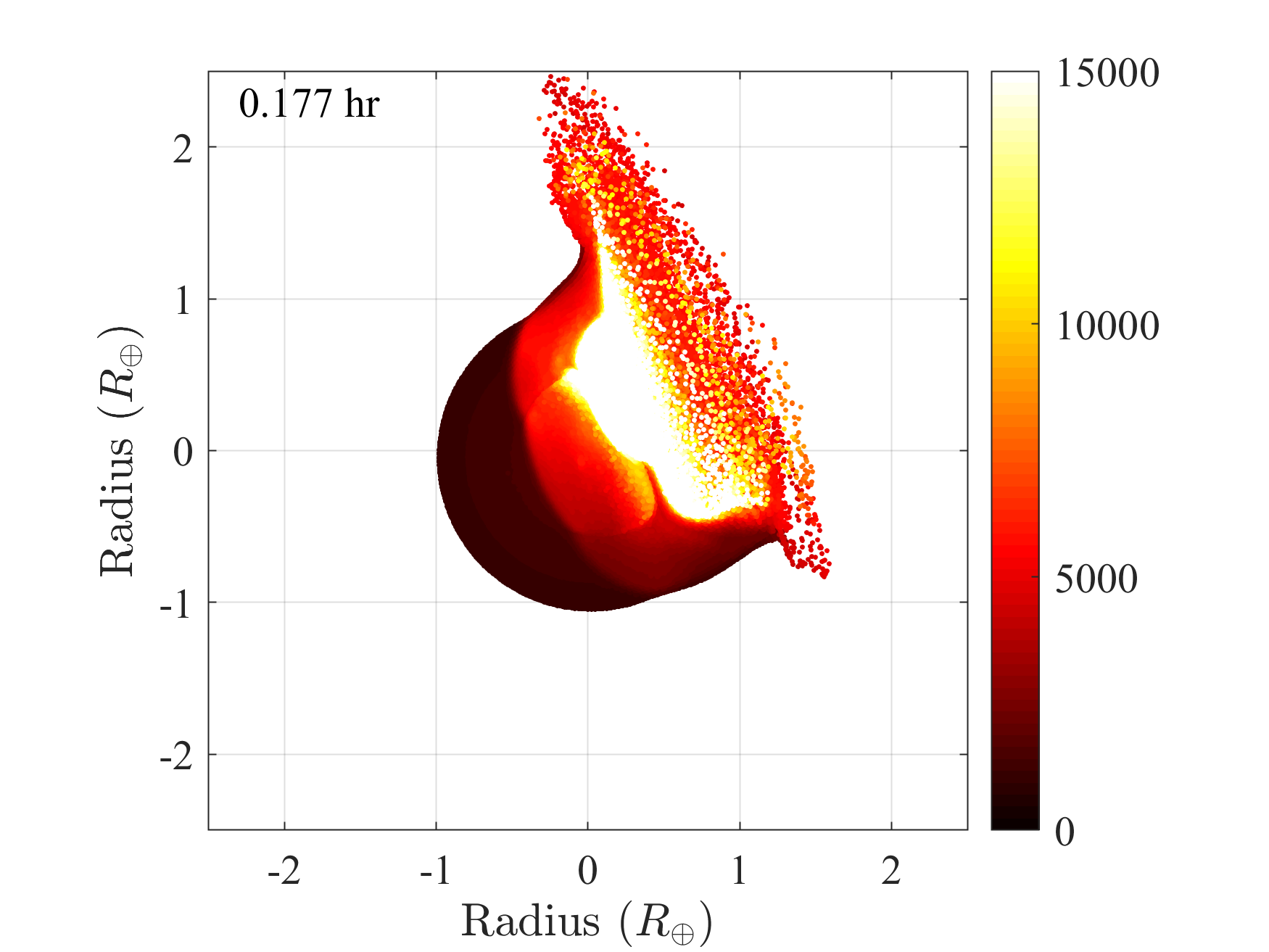}
\caption{Temperature distribution during the giant impact. All particles are projected onto the X-Y plane after being sorted by temperature. The color bar represents the temperature.
\label{fig:s2}}
\end{figure}

\begin{figure}[!ht]
\centering
\includegraphics[scale=1]{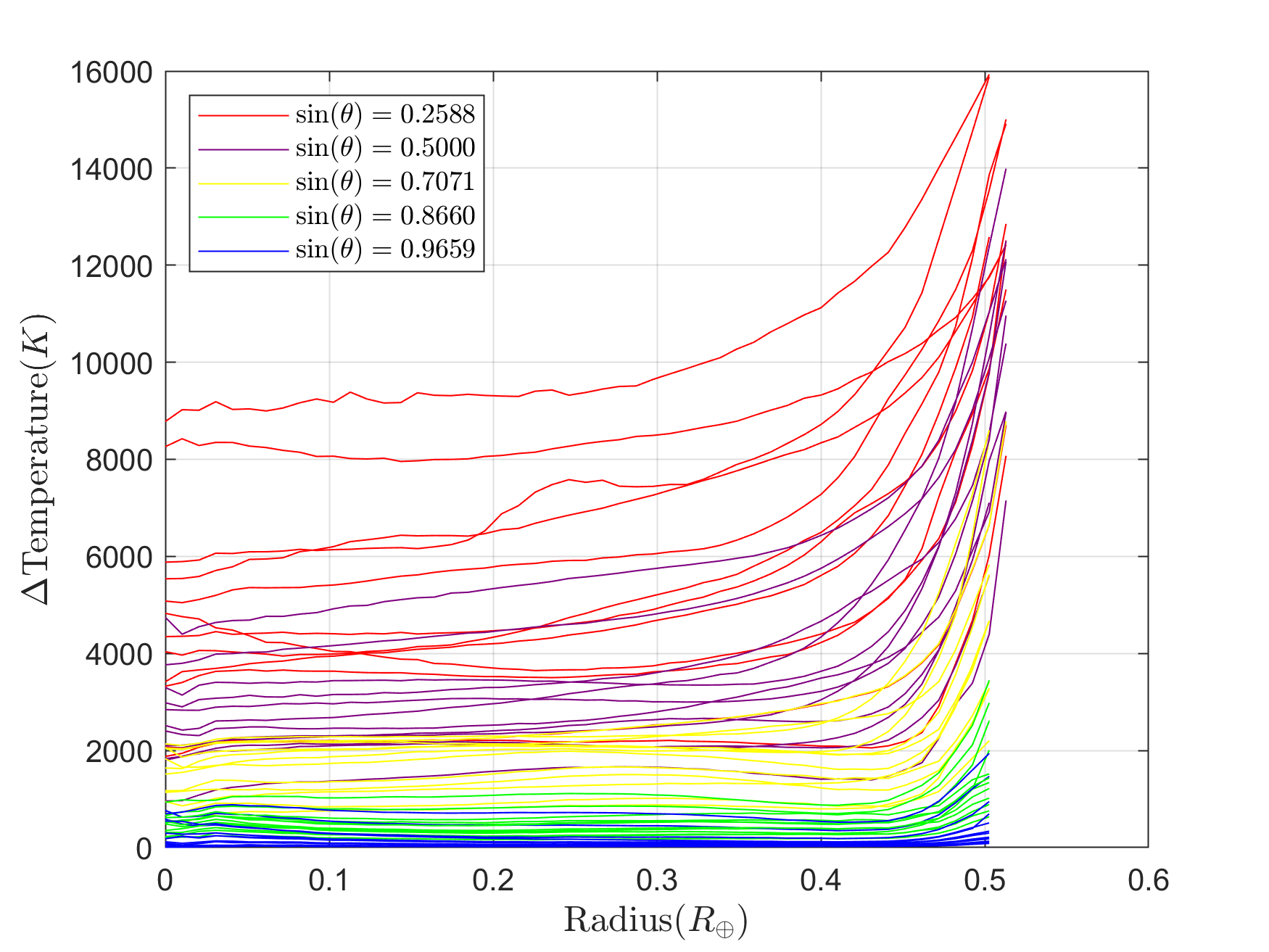}
\caption{The change in core temperature obtained from simulations against radius. The initial surface temperature is 2000~K, and different colors represent the impact angle. Each distinct line represents a different simulation, totaling 60 simulations. The region from the center of the core to the core-mantle boundary is segmented into 50 shells.
\label{fig:s3}}
\end{figure}

\begin{figure}[!ht]
\centering
\includegraphics[scale=1]{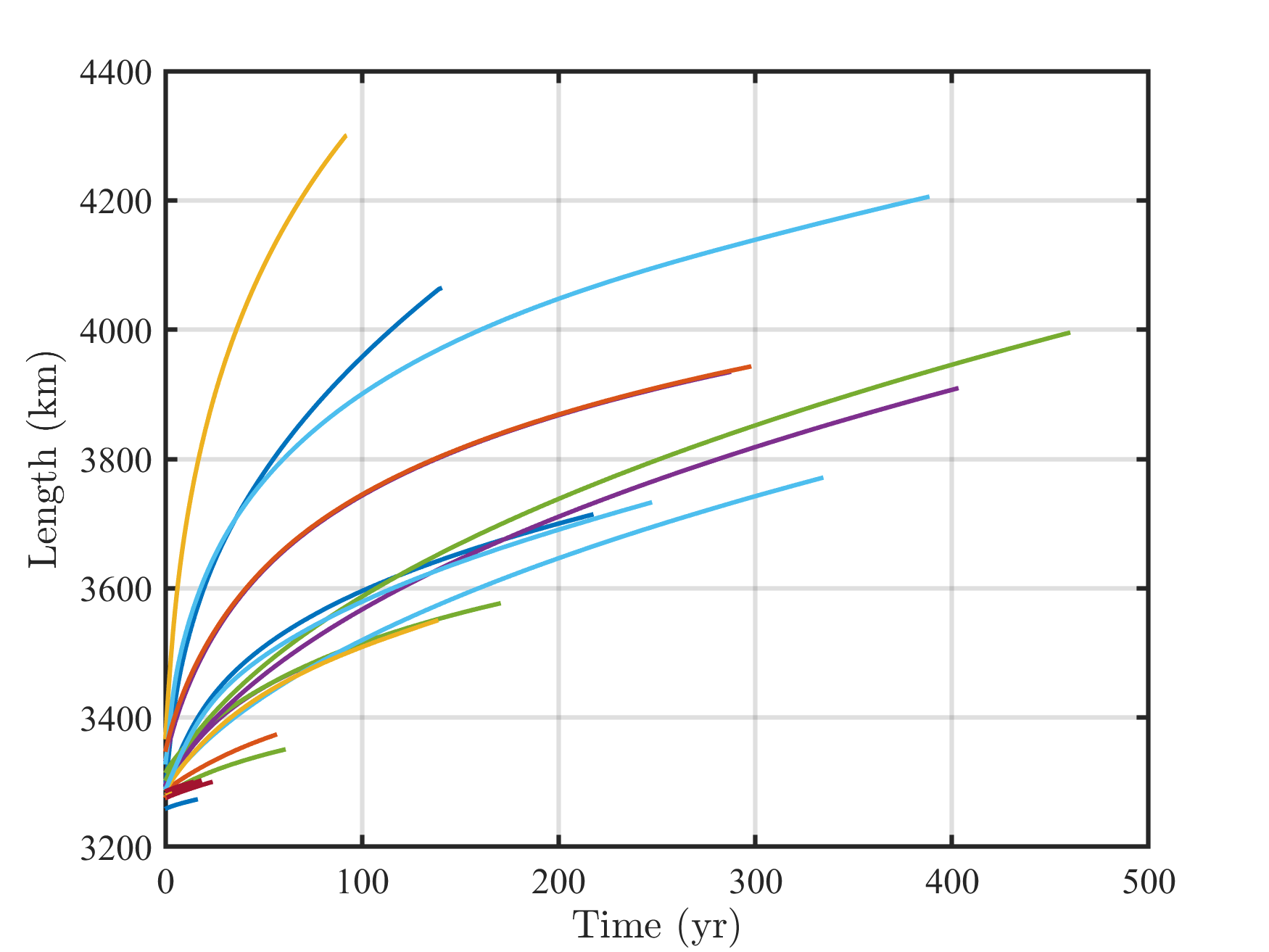}
\caption{Mantle melting evolution after different impacts. Colored curves indicate different initial viscosities or thermal conditions, illustrating the diversity in melting duration and evolution. The calculation adopts the viscosity model of Ke and Solomantov(2009); if other viscosity values are used, the melting duration could increase to several thousand years.
\label{fig:s4}}
\end{figure}

\begin{figure}[!ht]
\centering
\includegraphics[scale=1]{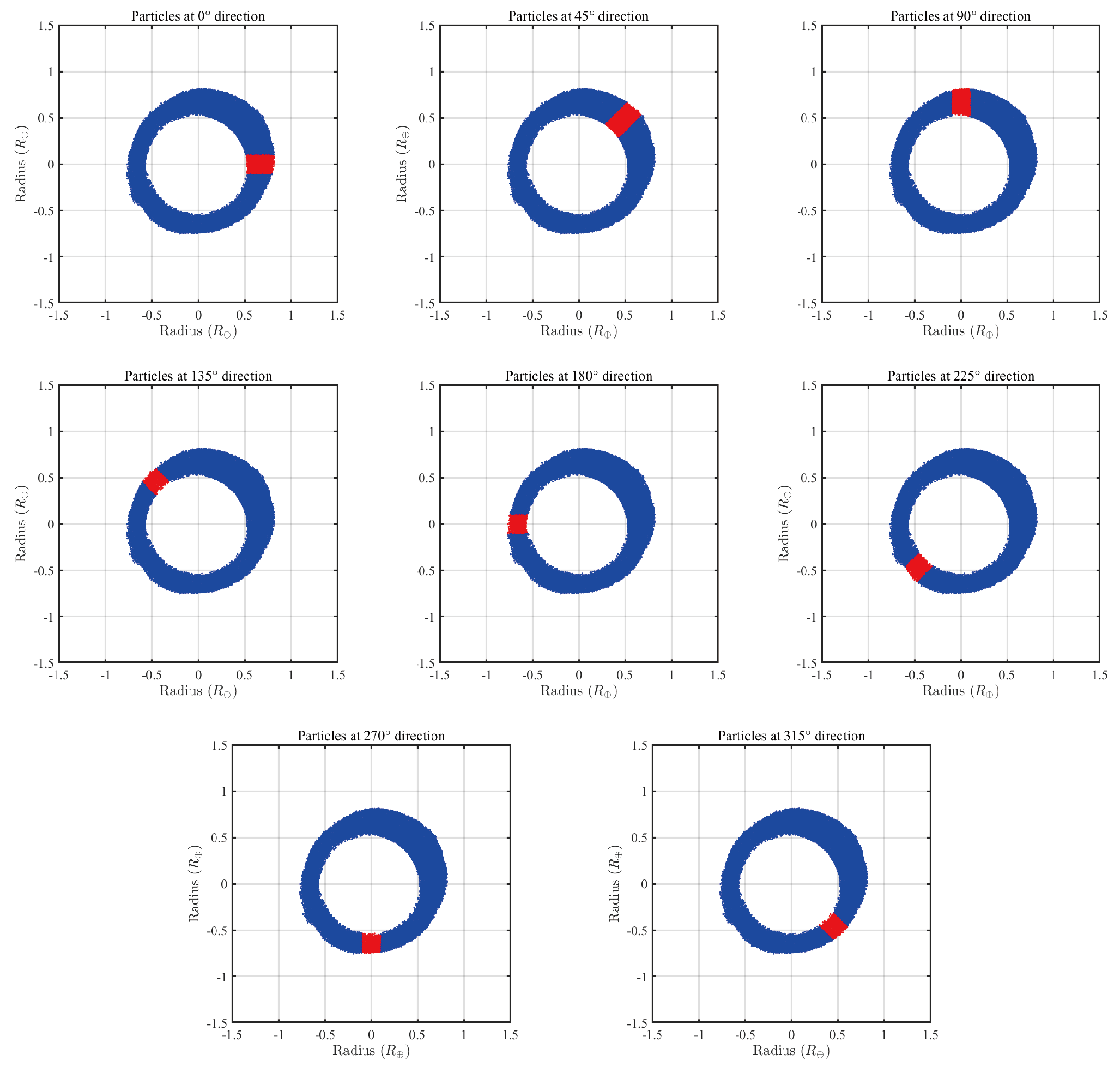}
\caption{
Schematic illustration of the procedure for estimating the average thickness of the solid lower mantle. The thickness was measured along eight directions (red sectors), and the mean value was taken as the representative thickness.
\label{fig:s5}}
\end{figure}

\end{document}